\documentclass[preprint,preprintnumbers, prd, floatfix,  superscriptaddress,nofootinbib]{revtex4-1}
\usepackage{graphicx}
\usepackage{epsfig}
\usepackage{bm}
\usepackage{amssymb}
\usepackage{float}
\usepackage{amsmath}
\usepackage{dcolumn}
\usepackage{cancel}
\usepackage[colorlinks]{hyperref}
\usepackage[usenames,dvipsnames]{color}
\usepackage{color}
\usepackage{epstopdf}
\usepackage{nccbbb}
\usepackage{ulem}
\hypersetup{
    breaklinks=false,
    pdfstartview={FitH},    colorlinks=true,       
    linkcolor=blue,          
    citecolor=red,        
    filecolor=magenta,      
    urlcolor=black,           
    anchorcolor=green,      
    linktocpage=true
}

\begin{document}

 \title{Running vacuum model in non-flat universe}

 \author{Chao-Qiang Geng}
\email{geng@phys.nthu.edu.tw}
\affiliation{School of Fundamental Physics and Mathematical Sciences\\Hangzhou Institute for Advanced Study, UCAS, Hangzhou 310024, China}
\affiliation{International Centre for Theoretical Physics Asia-Pacific, Beijing/Hangzhou, China}
\affiliation{Department of Physics, National Tsing Hua University,
Hsinchu, Taiwan 300}
 \affiliation{National Center for Theoretical Sciences, Hsinchu,
Taiwan 300}
 \author{Yan-Ting Hsu}
\email{ythsu@gapp.nthu.edu.tw}
\affiliation{Department of Physics, National Tsing Hua University,
Hsinchu, Taiwan 300}
\author{Lu Yin}
\email{yinlu@gapp.nthu.edu.tw}
\affiliation{School of Fundamental Physics and Mathematical Sciences\\Hangzhou Institute for Advanced Study, UCAS, Hangzhou 310024, China}
\affiliation{International Centre for Theoretical Physics Asia-Pacific, Beijing/Hangzhou, China}
\affiliation{Department of Physics, National Tsing Hua University,
Hsinchu, Taiwan 300}
\author{Kaituo Zhang}
\email{Corresponding author, ktzhang@ahnu.edu.cn  }
\affiliation{Department of Physics, Anhui Normal University, Wuhu, Anhui 241000, China}
\affiliation{Collaborative Innovation Center of Light Manipulations and Applications, Shandong Normal University, Jinan 250358, China}

\begin{abstract}
We investigate observational constraints on the running vacuum model (RVM) of $\Lambda=3\nu (H^{2}+K/a^2)+c_0$ in the spatially curved universe, where $\nu$ is the model parameter, $K$ corresponds to the spatial curvature constant, $a$ represents the scalar factor, and $c_{0}$ is a constant defined by the boundary conditions. We study the CMB power spectra with several sets of $\nu$ and $K$ in RVM. By fitting the cosmological data, we find that the best fitted $\chi^2$ value of RVM is slightly smaller than that of $\Lambda$CDM in the non-flat universe, along with the constraints of  $\nu\leq O(10^{-4})$ (68 $\%$ C.L.) and $|\Omega_K=-K/(aH)^2|\leq O(10^{-2})$ (95 $\%$ C.L.). In particular, our results favor the open universe in both  $\Lambda$CDM and RVM. In addition, we show that the cosmological constraints of $\Sigma m_{\nu}=0.256^{+0.224}_{-0.234}$  (RVM) and $\Sigma m_{\nu}=0.257^{+0.219}_{-0.234}$  ($\Lambda$CDM) at 95$\%$ C.L.  for the neutrino mass sum are relaxed in both models in the spatially curved universe.
 \end{abstract}


\maketitle
\section{Introduction}
Since the discovery that our universe has been expanding at an accelerated rate at recent time from the type Ia supernova data~\cite{Perlmutter:1998,Riess:1998,Perlmutter:1999},
many dark energy models have been proposed to explain these phenomena~\cite{Bronstein1933,Turner1997,Wetterich2002,Copeland:2006,Bamba2012,limiao}. 
  The simplest one is 
the $\Lambda$CDM model, in which $\Lambda$ represents the cosmological constant term. 
However, $\Lambda$CDM encounters some difficulties, mainly the ``fine tuning''~\cite{Weinberg:1989,Weinberg:1972} and ``coincidence''~\cite{Ostriker:1995,Wang:2000,ArkaniHamed:2000} problems.

The running vacuum model (RVM)~\cite{Ozer:1986,Carvalho:1991} has been introduced in order to solve the ``coincidence problem'', 
where the cosmological constant term is assumed to be varying with the Hubble parameter $H$.   
This model links the existence of dark energy to the theoretical mechanism of the quantum field,
which may trigger the primordial inflation scenario~\cite{Sola:2008}, and  fit with  the observational data better than $\Lambda$CDM~\cite{Geng20172}. 
In the literature, the spatially flat RVM has been extensively  investigated~\cite{EspanaBonet:2004,Sola:2013,Gomez-Valent:20151,Gomez-Valent:20152,Sola:2015,Gomez-Valent:20153,Geng2016,Sola2017,Sola:2019uum,Geng20171,Geng20172,Fritzsch:2017,Basilakos2018,Sola2018,Zhang2019,Geng:2020mga,Basilakos2019,Tsiapi2019}.

Recently, the Planck Legacy 2018 analysis by
Valentino, Melchiorri and Silk  in Ref.~\cite{Valentino2019} has suggested 
that the universe is  closed at 99 $\% $ C.L.~\cite{Handley2019}. 
 Similar conclusions have been obtained by Park and Ratra  in the context of non spatially-flat DE models obtained~\cite{Ratra20181,Ratra20182,Ratra20183,Ratra20191}. 
 Nevertheless, when the BAO data set is included together with the CMB, the evidence in favor of a non spatially-flat universe disappears completely~\cite{Aghanim:2018eyx,Valentino2019,Efstathiou2020}. These interesting results encourage us  to study RVM in a non-flat universe~\cite{Shapiro2003,Basilakos:2009} besides the flat one~\cite{Geng20172}. With the involvement of the non-zero spatial curvature, it is inevitable to encounter the degeneracies between curvature and other parameters. One of them is the famous ``geometrical degeneracy''~\cite{{Efstathiou:1998xx}, {Howlett:2012mh}} on CMB power spectra, caused by different sets of parameters that lead to same value of the angular diameter distance of the last scattering. 
On the other hand, when fitting with the observational data, the non-zero spatial curvature also broadens the constraints of the cosmological parameters~\cite{{Farooq:2013dra}}.

In this work, we concentrate on the running cosmological constant in the non-flat universe, $\Lambda=3\nu (H^{2}+ K/a^2)+c_0$, 
where $\nu$ and $c_{0}$ are the model parameters, $a$ represents the scalar factor and $K$ corresponds to the spatial curvature constant. 
 We first study the CMB power spectra in this non-flat RVM and discuss the degeneracy between $\nu$ and the  density parameter of curvature $\Omega_K=-K/(aH)^2$. We then constrain the cosmological parameters of 
both non-flat RVM and $\Lambda$CDM with the observational data by using the Markov chain Monte Carlo (MCMC) method, and compare the results with those  in the flat universe. The effectiveness of RVM versus $\Lambda$CDM in the  non-flat universe is also tested based on the minimal $\chi^2$ values.

This paper is organized as follows. 
In Sec.~\ref{sec:The Curved RVM Model}, we introduce the non-flat running vacuum model and derive the background evolution equations. We compare the CMB power spectra of RVM in the non-flat universe along with the Planck 2018 data, and show the constraints  of the cosmological parameters in Sec.~\ref{sec:numerical}. Our  conclusions is presented in Sec.~\ref{sec:conclusion}.

\section{ Evolution of RVM  in curved Universe}\label{sec:The Curved RVM Model}

 
We start with the Einstein field equation of RVM, given by
 \begin{equation}
\label{eq:EisteinEqu}
 R_{\alpha\beta}-\frac{1}{2} g_{\alpha\beta} R+g_{\alpha\beta} \Lambda= \kappa^2 T_{\alpha\beta},
 \end{equation}
where $\kappa^2=8\pi G$ is set to be 1  for simplicity, $R=g^{\alpha\beta}R_{\alpha\beta}$ represents the Ricci scalar, 
 $T_{\alpha\beta}$ stands for the energy-momentum tensor for matter and radiation, 
 and $\Lambda$ corresponds to the dynamical cosmological constant.

The spatially isotropic and homogeneous universe can be described by the Robertson-Walker metric:
\begin{equation}
\label{eq:RWmetric}
ds^{2}=-dt^{2}+a^2(t)\left\{\frac{dr^2}{1-Kr^2} +r^2d\theta^2+r^2\sin^2\theta d\phi^2 \right\}\,,
\end{equation}
where $a$ is the scale factor, while $K$ is a constant describe the spatial curvature with $K=1,0,-1$ corresponding to the closed, flat, and open universe, respectively. Then, the Friedmann equations can be expressed as
\begin{eqnarray}
\label{eq:friedmann1}
&& H^{2}=\frac{1}{3}(\rho_m + \rho_r  + \rho_{\Lambda})-\frac{K}{a^2} \,, \\
\label{eq:friedmann2}
&& \dot{H}=- \frac{1}{2} (\rho_m + \rho_r + \rho_{\Lambda} + P_m+ P_r + P_{\Lambda})+\frac{K}{a^2}, \,
\end{eqnarray}
where $\rho_{m, r, \Lambda}$ ($P_{m, r, \Lambda}$) are the energy densities (pressures) of matter, radiation and dark energy, respectively, and $H=da/(adt)$ represents the Hubble parameter. 
We note that $\rho_{\Lambda}=\kappa^{-2} \Lambda$, and the density parameters are parameters are given by
	\begin{eqnarray}
	\label{eq:omega}
	&&\Omega_{m, r}=\frac{\rho_{m, r}}{3H^2} , \\
	&&\Omega_{\Lambda}=\frac{\Lambda}{3H^2}   , \\
	&&\Omega_K=-\frac{K}{a^2 H^2}. \,
	\end{eqnarray}

In the non-flat universe, the running cosmological constant term is set to be
\begin{equation}
\label{eq:Lambda}
\Lambda=3\nu H^{2}+3\nu \frac{K}{a^2}+c_0\,,
\end{equation}
where $\nu$ is a non-negative model parameter to ensure that the energy density of dark energy is positive in the early universe, $c_0$ is given by
 $c_0 = -3\nu ( H_0^2+K)+\Lambda_0$ with $H_0$ and $\Lambda_0$ the present values of the Hubble parameter and
 cosmological constant, respectively. The model becomes to be  $\Lambda$CDM when $\nu = 0$. 
The corresponding equations of state in this model can be defined as
 \begin{equation}
 \label{eq:eos}
 w_{m, r, \Lambda}=\frac{P_{m, r, \Lambda}}{\rho_{m, r, \Lambda}}= 0, \frac{1}{3}, -1 \,.
 \end{equation}
For the energy transformations from dark energy to matter and radiation, the modified continuity equations are given by
\begin{eqnarray}
\label{eq:continuity}
&& \dot{\rho}_{m, r}+3 H(1+w_{m, r})\rho_{m, r} = Q_{m, r} \,, \\
&& \dot{\rho}_\Lambda+3 H(1+w_\Lambda)\rho_\Lambda = - Q \,,
\end{eqnarray}
with $Q_{m}+Q_{r}=Q$, $Q_{m, r}$ can be  
written as
\begin{eqnarray}
\label{eq:coupling}
Q_{m, r}=-\frac{\dot{\rho}_{\Lambda}(\rho_{m, r}+P_{m, r})}{\rho_{m}+\rho_{r}+P_m+P_r}=3\nu H(1+w_{m, r})\rho_{m, r}\,.
\end{eqnarray}
By combining Eqs.~({\ref{eq:friedmann1}})-({\ref{eq:coupling}}), we derive the energy densities as functions of the scale factor:
\begin{eqnarray}
\label{eq:rhom}
&&\rho_{m}(a)=\rho_{m}^{(0)}a^{-3\xi} , \\
\label{eq:rhor}
&&\rho_{r}(a)=\rho_{r}^{(0)}a^{-4\xi} , \\
\label{eq:rhode}
&&\rho_{\Lambda}(a)=\rho_{\Lambda}^{(0)}+\left(\xi^{-1}-1\right)\left [\rho_m^{(0)}\left(a^{-3\xi}-1\right)+\rho_r^{(0)}\left(a^{-4\xi}-1\right)\right]  ,\
\end{eqnarray}
where $\rho_{m,r.\Lambda}^{(0)}$ are current values and $\xi=(1-\nu)$.
Consequently, the Friedmann equation defined in Eq.~({\ref{eq:friedmann1}}) can be  rewritten as
\begin{eqnarray}
\label{eq:friedmann3}
H^{2}=H_{0}^{2}\left\{\frac{1}{\xi}\left[\Omega_{m}^{0}\left(a^{-3\xi}-1\right)+\Omega_{r}^{0}\left(a^{-4\xi}-1\right)\right]+\Omega_{K}^{0}\left(a^{-2}-1\right)+1\right\} ,\
\end{eqnarray}
where $\Omega_{m,r,K}^{0}$ are current values of density parameters.

\section{Numerical Calculations}\label{sec:numerical}

To study the degeneracy between the cosmological parameters, we first modify the {\bf CAMB}~\cite{Lewis:1999bs} program 
to generate theoretical CMB power spectra for  both models of RVM and $\Lambda$CDM. 
The results are presented in Sec.~\ref{sec:A}. We then use the {\bf CosmoMC} package~\cite{Lewis:2002ah}, which is a Markov Chain Monte Carlo (MCMC)
 engine exploring the cosmological parameter space, to constrain RVM and $\Lambda$CDM from the observational data. For simplification, we take 
  $\Omega_K$ afterward to represent the density parameter of curvature at the present time except those specifically indicated.

\subsection{CMB power spectra of the models}\label{sec:A}
There is ``geometrical degeneracy'' between curvature and other parameters for CMB power spectra. 
To see this effect, we compare CMB power spectra of RVM and $\Lambda$CDM with different $\nu$ and $\Omega_K$ along with the observational data
from Planck 2018~\cite{Aghanim:2018eyx}. From the previous studies in the literature~\cite{Zhang2019,Geng20172,Geng:2020mga} 
 with  
 $0 \leq \nu \leq O(10^{-3})$ in RVM for the flat universe and the result of  $-0.007 \geq \Omega_K \geq -0.095$ at 99 $\%$ C.L. in Ref.~\cite{Valentino2019}, we choose $0 \leq \nu < 0.01$ and $0 \geq \Omega_K \geq -0.01$ to see the degeneracy between $\nu$ and $\Omega_K$ on CMB power spectra. 
 Furthermore, the $\Lambda$CDM model is recovered when $\nu=0$ and $\Omega_K=0$  in Eq.~\ref{eq:Lambda}.

\begin{figure}[h]
	\centering
\includegraphics[angle=-90,width=79mm]{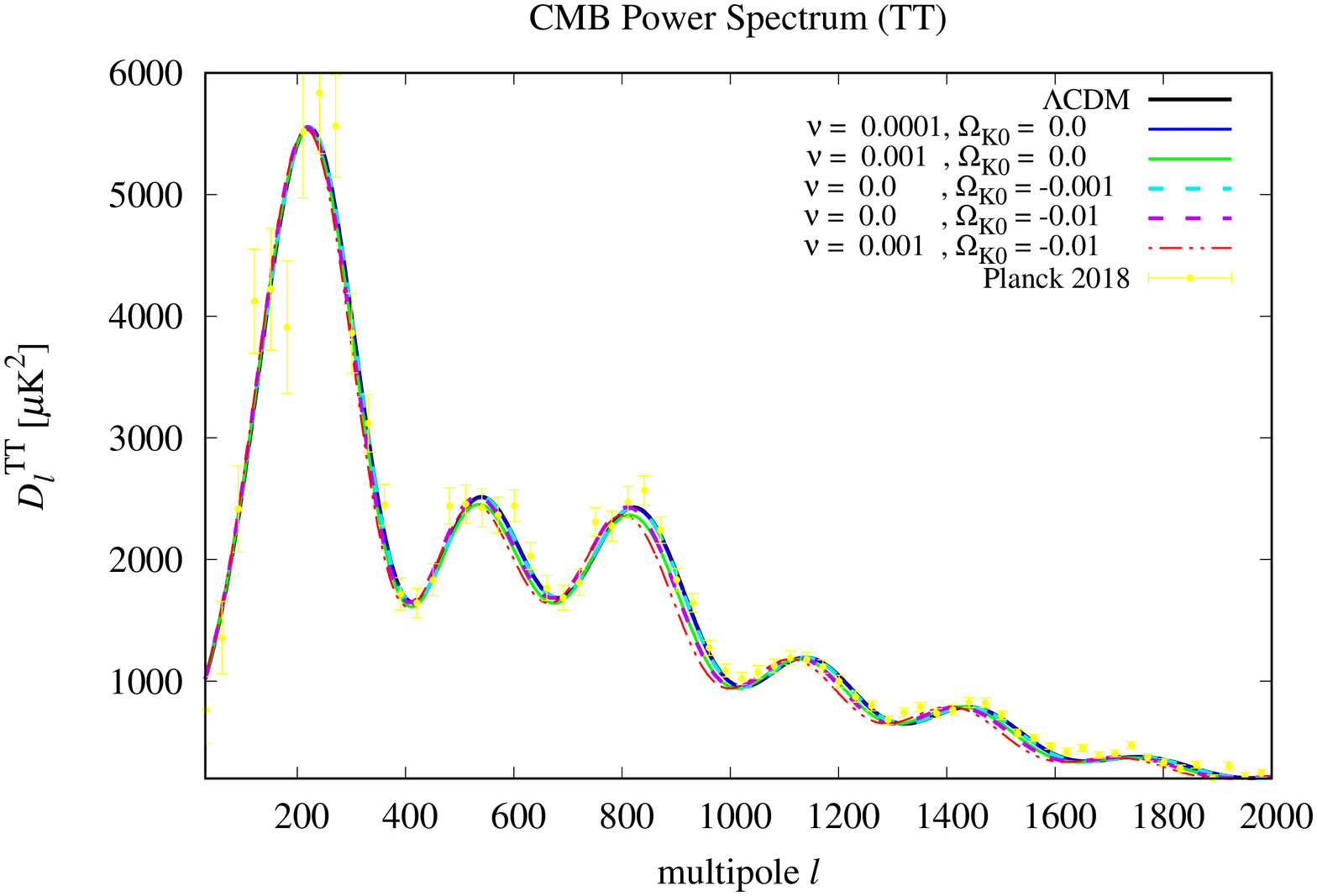}
\includegraphics[angle=-90,width=79mm]{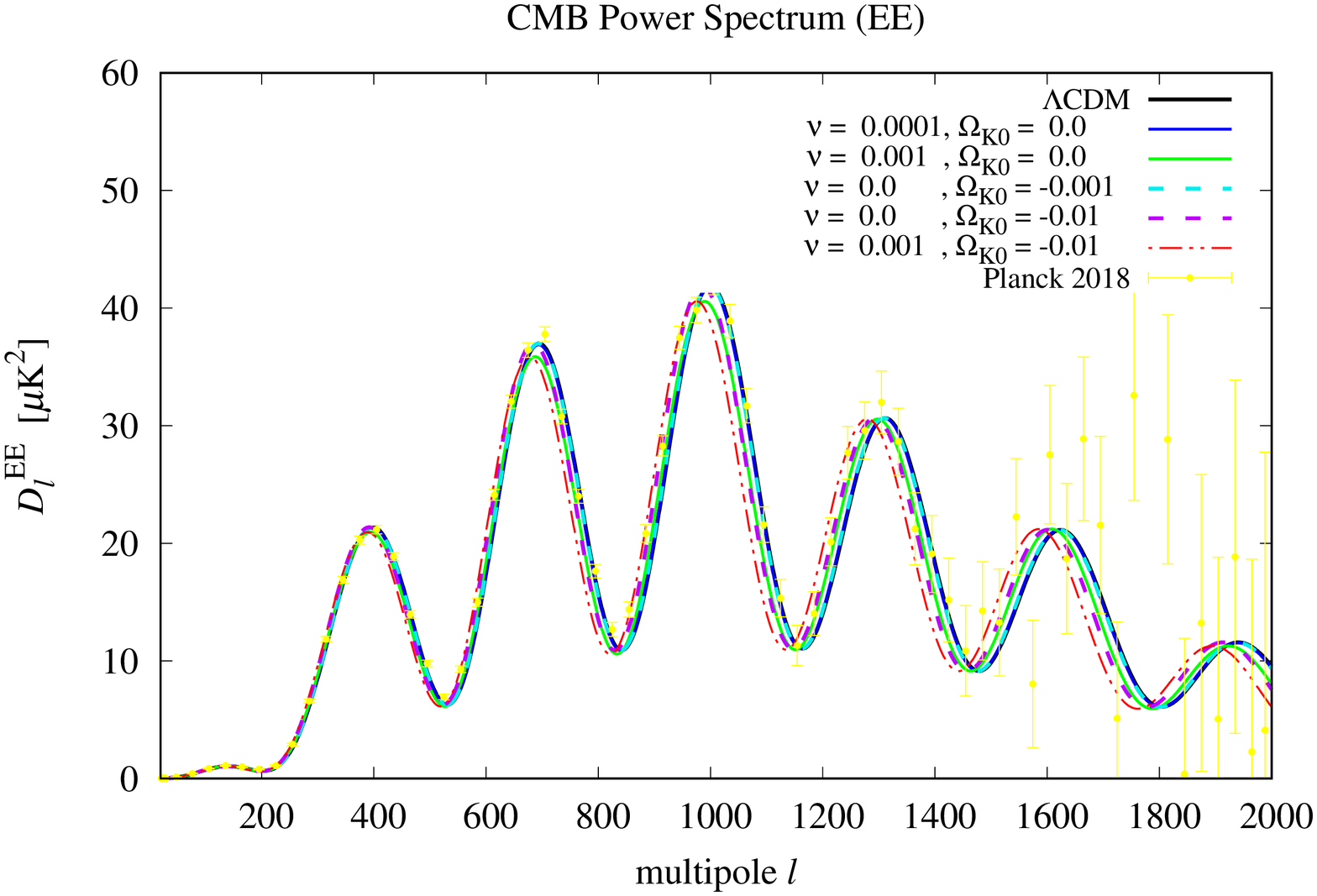}
\includegraphics[angle=-90,width=79mm]{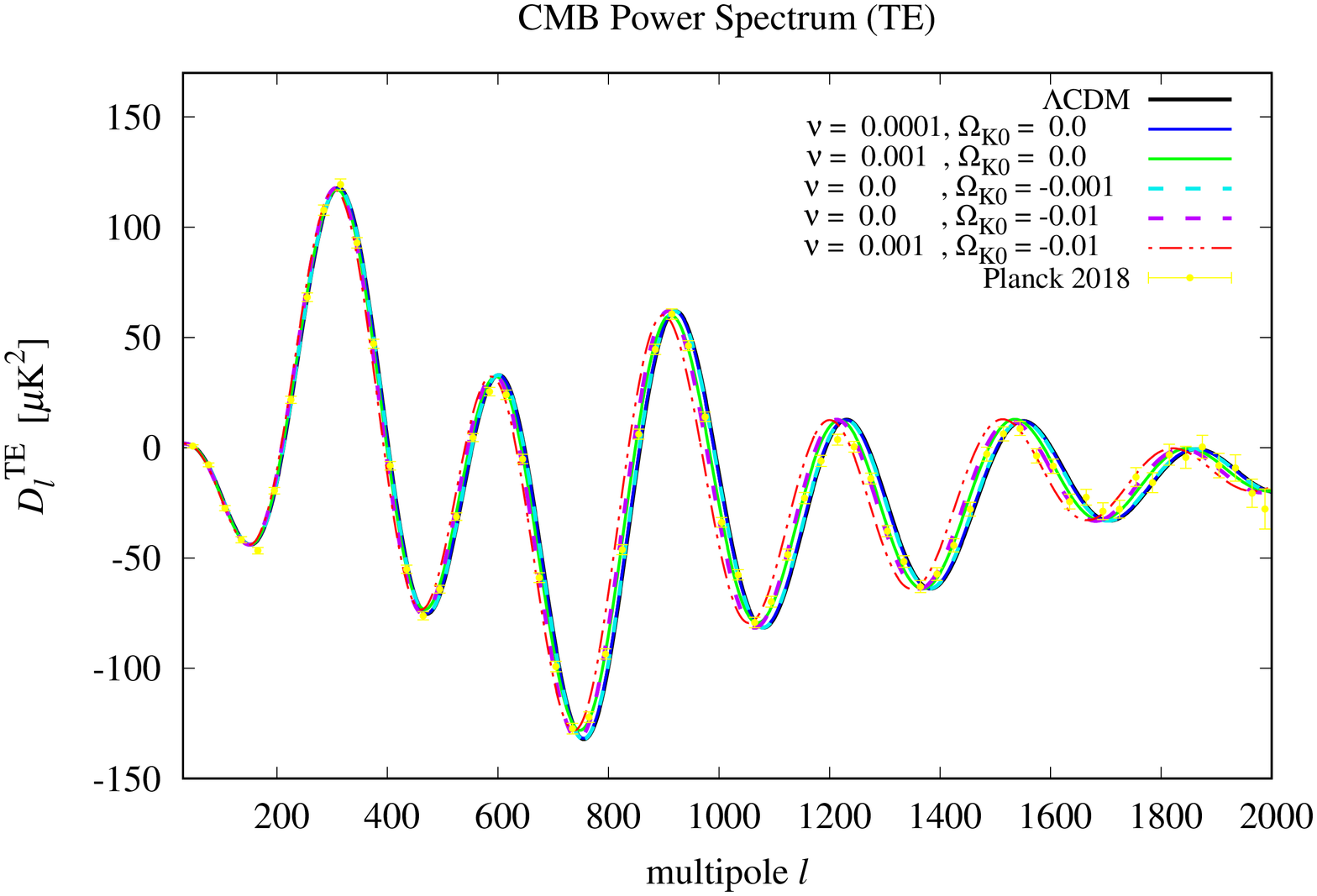}\\
	\caption{Power spectra of CMB TT, EE and TE  for RVM and $\Lambda$CDM in the flat and non-flat universe along with the Planck 2018 data.} 
	\label{fg:TT}
\end{figure}

\begin{figure}[h]
	\centering
\includegraphics[angle=-90,width=79mm]{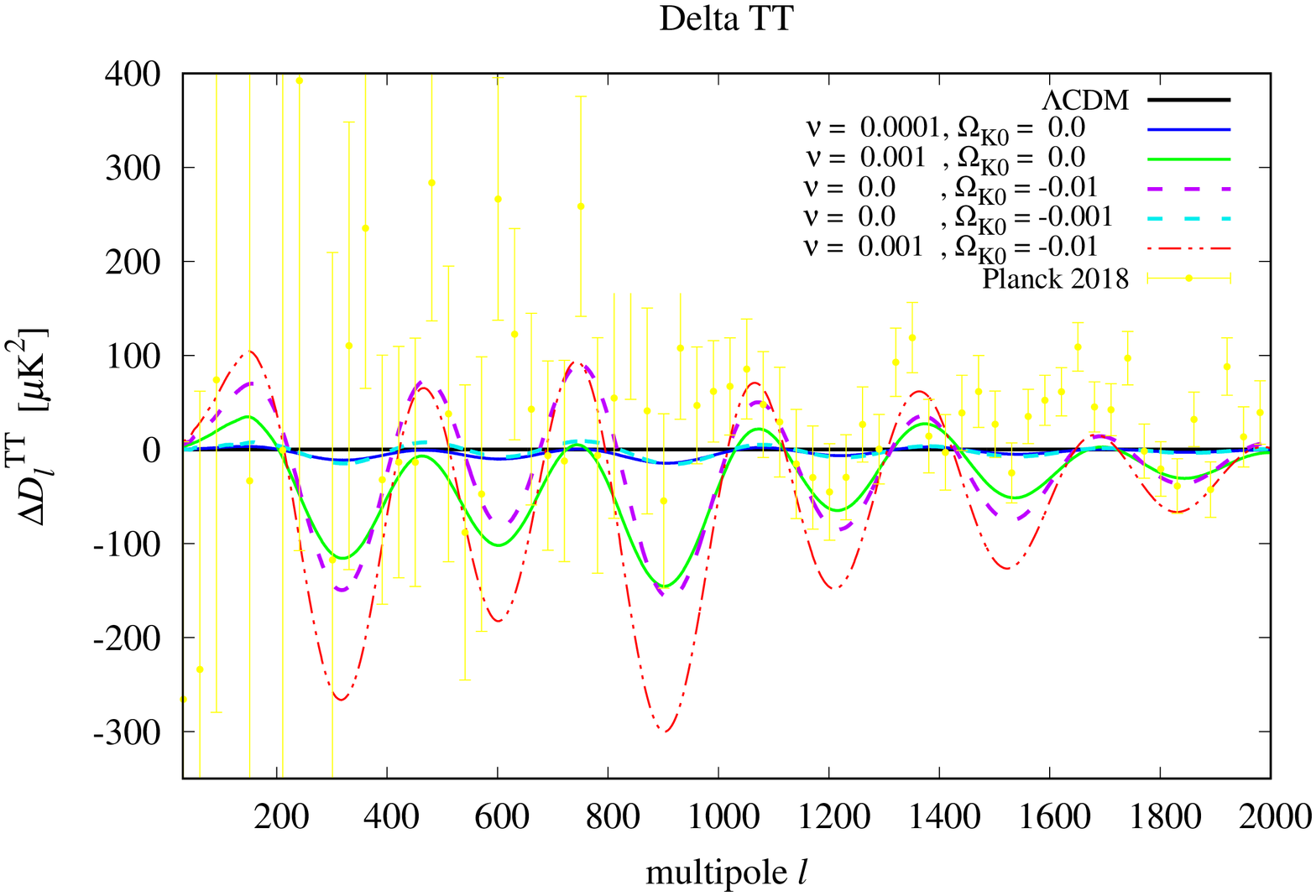}
\includegraphics[angle=-90,width=79mm]{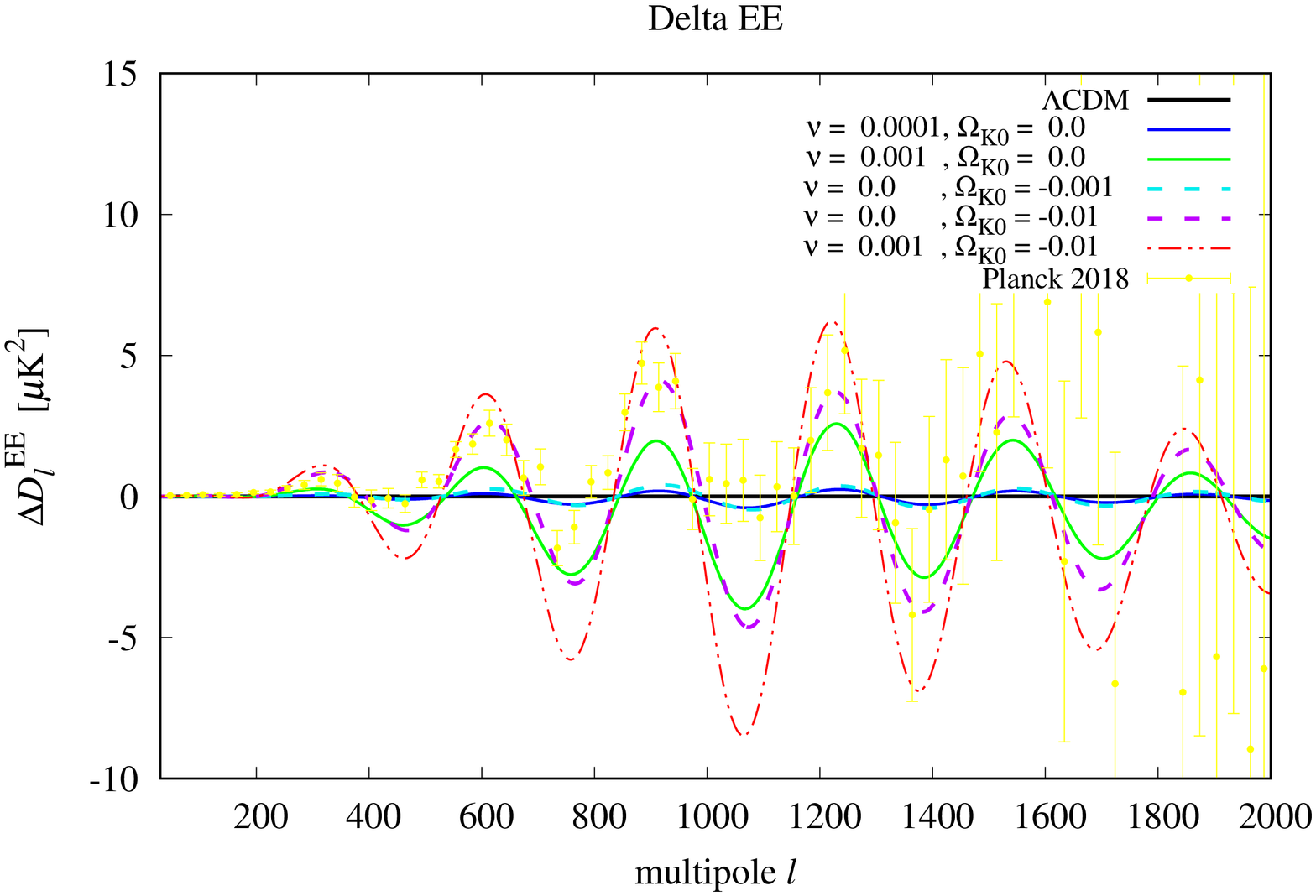}
\includegraphics[angle=-90,width=79mm]{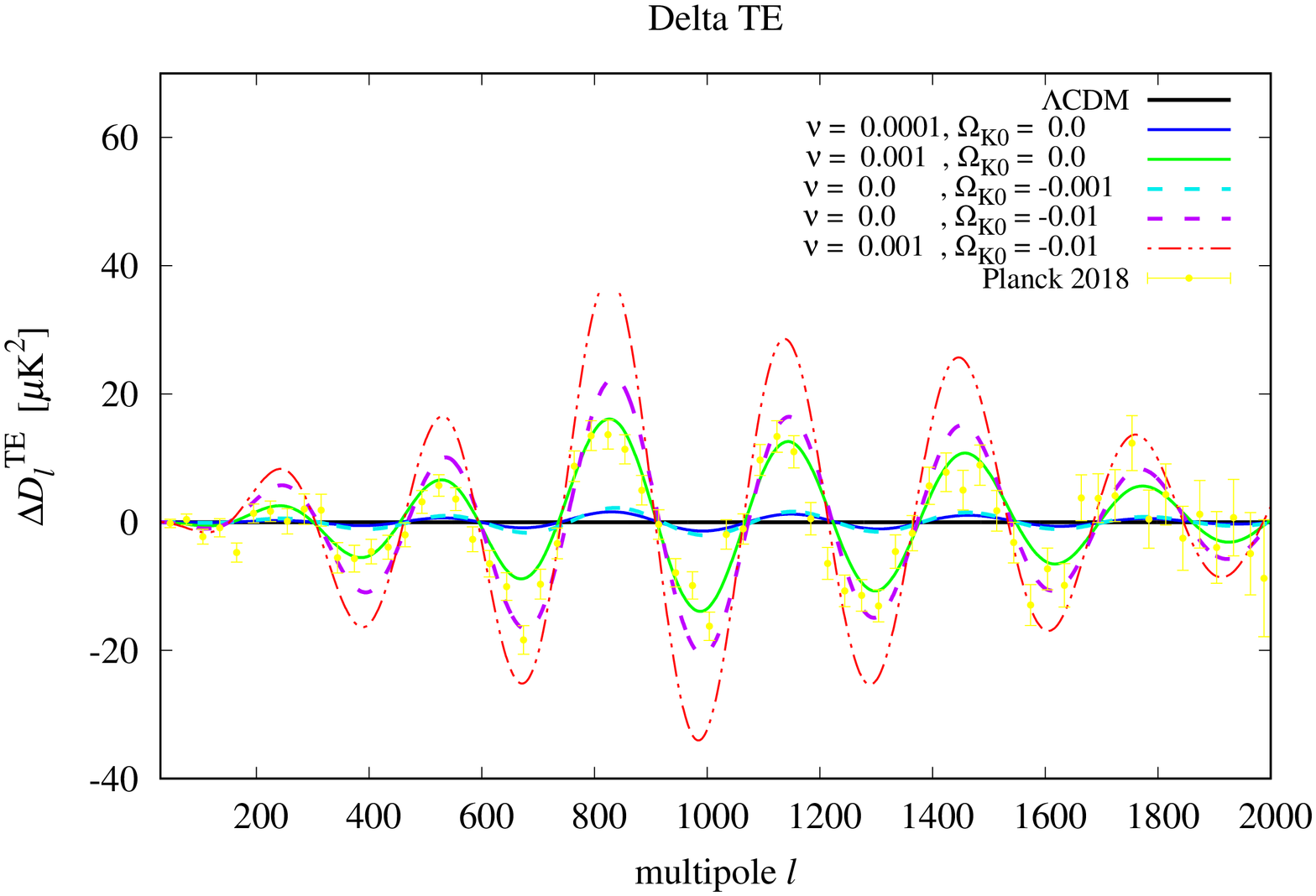}
	\caption{ Residuals of $\Delta D_{\ell}^{TT}$, $\Delta D_{\ell}^{EE}$ and  $\Delta D_{\ell}^{TE}$ in RVM with respect to $\Lambda$CDM for CMB power spectra, respectively, along with the observational data  from Planck 2018.} 
	\label{fg:diffTT}
\end{figure}

In Fig.~\ref{fg:TT}, we present the CMB power spectra for the TT, EE and TE modes from the {\bf CAMB} package.
It can be seen that $0\leq\nu\leq O(10^{-3})$ (solid lines) and $0 \geq \Omega_K \geq -O(10^{-2})$ (dashed lines) fit well with the  data from Planck 2018. 
The residues with respect to $\Lambda$CDM are plotted in Fig.~\ref{fg:diffTT}. 
We find that
the geometrical degeneracy with $(\nu, \Omega_K)=(0.001,0)$ (green solid line) and $(0.0, -0.01)$ (purple dashed line)
has the similar results on CMB power spectra. However, only  $\nu$ can cause strong suppressions on the TT mode spectra when $\nu > 0$. 
In addition, the effects of $\nu$ and $K$  show additive property on CMB power spectra (red dash-dotted line).

\subsection{Global fitting}

\begin{table}[!hbp]
	\caption{ data points of $f\sigma_8$}
	\begin{tabular}{|c|c|c|c||c|c|c|c||c|c|c|c|}
		\hline
		\ & $z$ & $f\sigma_8$ & Ref. & \ & $z$ & $f\sigma_8$ & Ref. & \ & $z$ & $f\sigma_8$ & Ref.  \\
		\hline
		~$1$~ & $1.36$ & $0.482 \pm 0.116$ & \cite{Okada:2015vfa} &
		~$10$~ & $0.59$ & $0.488 \pm 0.06$ & \cite{Chuang:2013wga} &
		~$19$~ & $0.35$ & $0.440 \pm 0.05$ & \cite{Song:2008qt, Tegmark:2006az}\\
		\hline
		$2$ & $0.8$ & $0.470 \pm 0.08$ & \cite{delaTorre:2013rpa} &
		$11$ & $0.57$ & $0.444 \pm 0.038$ & \cite{Gil-Marin:2015sqa} &
		$20$ & $0.32$ & $0.394 \pm 0.062$ & \cite{Gil-Marin:2015sqa} \\
		\hline
		$3$ & $0.78$ & $0.38 \pm 0.04$ & \cite{Blake:2011rj} &
		$12$ & $0.51$ & $0.452 \pm 0.057$ & \cite{Satpathy:2016tct} &
		$21$ & $0.3$ & $0.407 \pm 0.055$ & \cite{Tojeiro:2012rp} \\
		\hline
		$4$ & $0.77$ & $0.490 \pm 0.18$ & \cite{Song:2008qt, Guzzo:2008ac} &
		$13$ & $0.5$ & $0.427 \pm 0.043$ & \cite{Tojeiro:2012rp} &
		$22$ & $0.25$ & $0.351 \pm 0.058$ & \cite{Samushia:2011cs}\\
		\hline
		$5$ & $0.73$ & $0.437 \pm 0.072$ & \cite{Blake:2012pj} &
		$14$ & $0.44$ & $0.413 \pm 0.080$ & \cite{Blake:2012pj}&
		$23$ & $0.22$ & $0.42 \pm 0.07$ & \cite{Blake:2011rj}\\
		\hline
		$6$ & $0.61$ & $0.457 \pm 0.052$ & \cite{Satpathy:2016tct} &
		$15$ & $0.41$ & $0.45 \pm 0.04$ & \cite{Blake:2011rj} &
		$24$ & $0.17$ & $0.51 \pm 0.06$ & \cite{Song:2008qt, Percival:2004fs} \\
		\hline
		$7$ & $0.60$ & $0.390 \pm 0.063$ & \cite{Blake:2012pj} &
		$16$ & $0.4$ & $0.419 \pm 0.041$ & \cite{Tojeiro:2012rp} &
		$25$ & $0.15$ & $0.49 \pm 0.15$ & \cite{Howlett:2014opa} \\
		\hline
		$8$ & $0.6$ & $0.433 \pm 0.067$ & \cite{Tojeiro:2012rp} &
		$17$ & $0.38$  & $0.430 \pm 0.054$ & \cite{Satpathy:2016tct} &
		$26$ & $0.067$ & $0.423 \pm 0.055$ & \cite{Beutler:2012px} \\
		\hline
		$9$ & $0.60$ & $0.43 \pm 0.04$ & \cite{Blake:2011rj} &
		$18$ & $0.37$ & $0.460 \pm 0.038$ & \cite{Samushia:2011cs} &
		$27$ & $0.02$ & $0.36 \pm 0.04$ & \cite{Hudson:2012gt} \\
		\hline
	\end{tabular}
	\label{tab:2}
\end{table}

\begin{table}[ht]
	\begin{center}
		\caption{ Priors for cosmological parameters with the non-flat RVM of $\Lambda=3\nu (H^{2}+ K/a^2)+c_0$ }
		\begin{tabular}{|c||c|} \hline
			Parameter & Prior
			\\ \hline
			RVM parameter $\nu$& $0.0 \leq \nu \leq 3.0\times10^{-4}$
			\\ \hline
			Curvature parameter $\Omega_K$& $-0.25 \leq \Omega_K \leq 0.2$
			\\ \hline
			Baryon density & $0.5 \leq 100\Omega_bh^2 \leq 10$
			\\ \hline
			CDM density & $0.1 \leq 100\Omega_ch^2 \leq 99$
			\\ \hline
			Optical depth & $0.01 \leq \tau \leq 0.8$
			\\ \hline
			Neutrino mass sum& $0 \leq \Sigma m_{\nu} \leq 2$~eV
			\\ \hline
			$\frac{\mathrm{Sound \ horizon}}{\mathrm{Angular \ diameter \ distance}}$  & $0.5 \leq 100 \theta_{MC} \leq 10$
			\\ \hline
			Scalar power spectrum amplitude & $2 \leq \ln \left( 10^{10} A_s \right) \leq 4$
			\\ \hline
			Spectral index & $0.8 \leq n_s \leq 1.2$
			\\ \hline
		\end{tabular}
		\label{tab:prior}
	\end{center}
\end{table}

\begin{figure}[h]
	\centering
	\includegraphics[width=0.96 \linewidth]{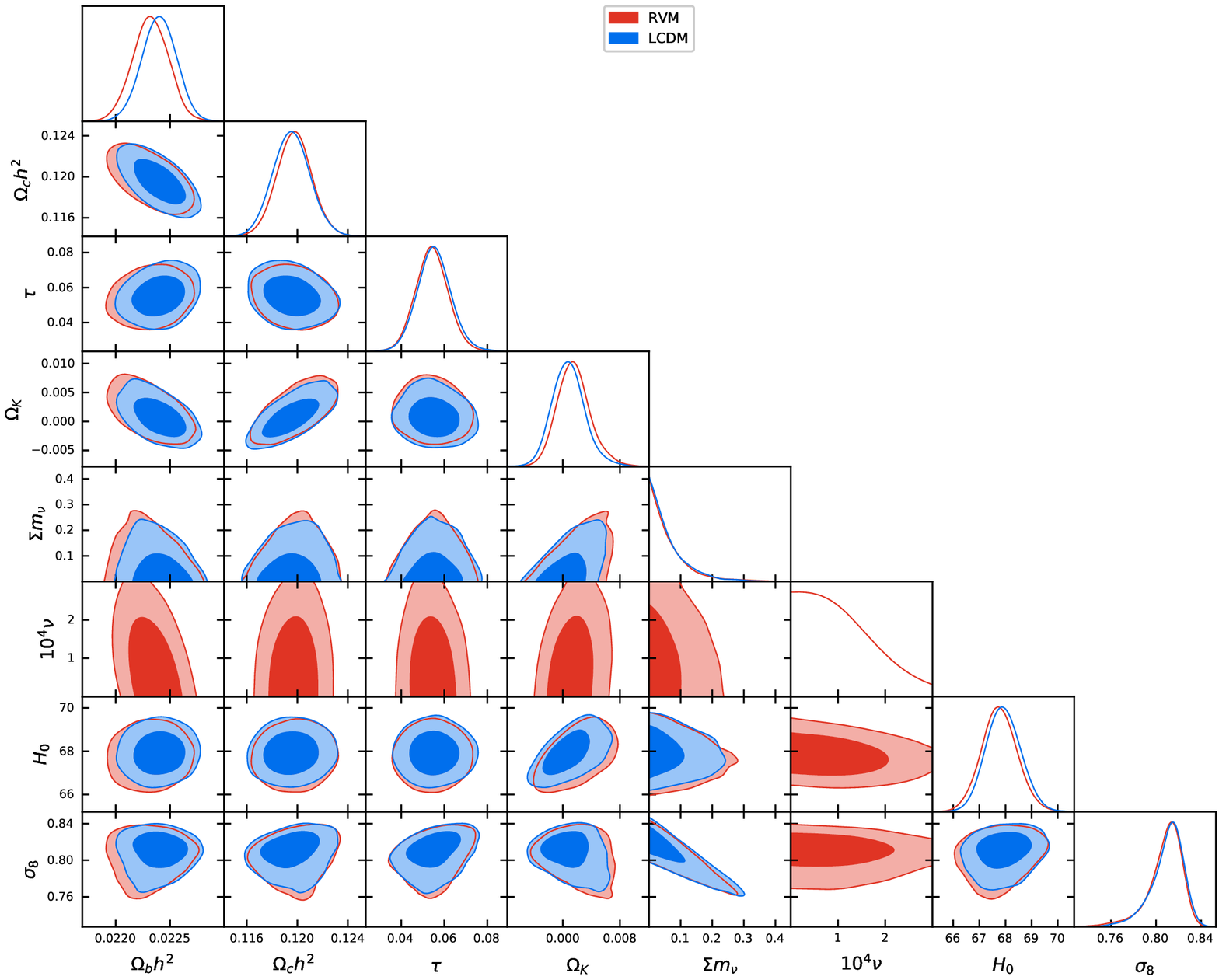}
	\caption{One and two-dimensional distributions of $\Omega_b h^2$, $\Omega_c h^2$, $\tau$, $\Omega_K$, $\sum m_\nu$, $10^4 \nu$, $H_0$, $\sigma_8$ for RVM and $\Lambda$CDM in the non-flat universe with the combined data of CMB+BAO+SN, where the contour lines represent 68$\%$~ and 95$\%$~ C.L., respectively.}
	\label{fig:2018cmbbaosn}
\end{figure}
\begin{figure}[h]
	\centering
	\includegraphics[width=0.96 \linewidth]{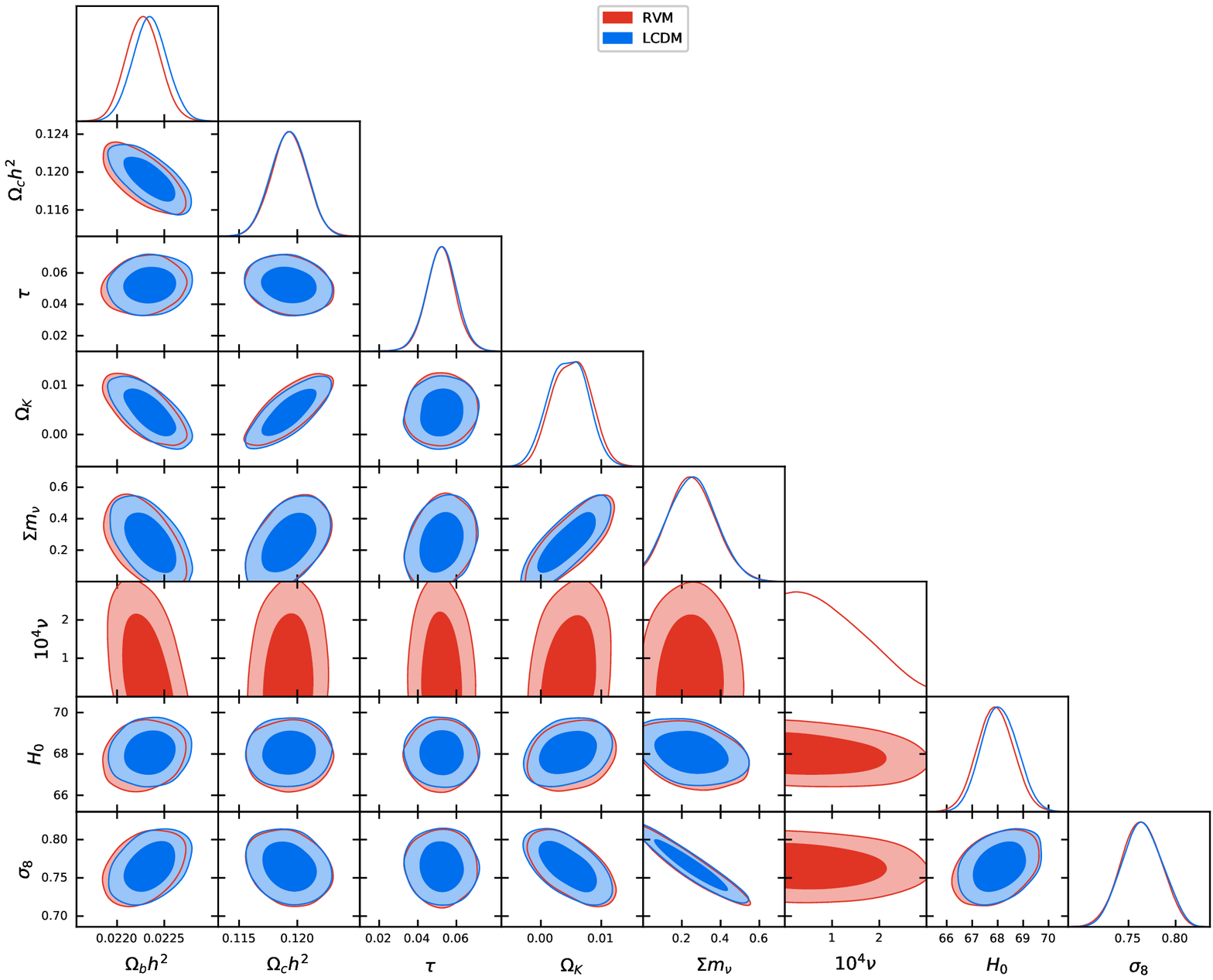}
	\caption{One and two-dimensional distributions of $\Omega_b h^2$, $\Omega_c h^2$, $\tau$, $\Omega_K$, $\sum m_\nu$, $10^4 \nu$, $H_0$, $\sigma_8$ for RVM and $\Lambda$CDM in the non-flat universe with the combined data of CMB+BAO+SN+WL+$f\sigma_8$, where the contour lines represent 68$\%$~ and 95$\%$~ C.L., respectively.}
	\label{fig:2018cmbbaosnwlfsigma8}
\end{figure}

\begin{table}[h]
	\newcommand{\tabincell}[2]{\begin{tabular}{@{}#1@{}}#2\end{tabular}}	
	\begin{center}
		\caption{Fitting results for  RVM and $\Lambda$CDM in the non-flat universe, where the cosmological parameters and $\nu$ are given at 95$\%$
		 and 68$\%$ C.L., respectively.}
		\begin{tabular} {|c|c|c|c|c|}
			\hline
			Parameter &
			\multicolumn{2}{|c|}{\tabincell{c}{
			CMB+BAO+SN
			\\}
		        } 
			&
			\multicolumn{2}{|c|}{\tabincell{c}{
			CMB+BAO+SN\\+WL+$f\sigma_8$
			}}
			\\
			\hline
			Model &RVM&$\Lambda$CDM&RVM&$\Lambda$CDM\\
			\hline			
			{\boldmath$100\Omega_b h^2   $} & 
			$2.23 \pm 0.03 $& 
			$2.24 \pm 0.03 $&
			$2.23 ^{+0.04}_{-0.03}$& 
			$2.23 \pm 0.04$\\
			
			{\boldmath$100\Omega_c h^2   $} &
			$12.0 \pm 0.3 $&
			$12.0 \pm 0.3 $&
			$11.9 \pm 0.3 $&
			$11.9 \pm 0.3 $\\
			
			{\boldmath$100\tau           $} & 
			$5.43 ^{+1.52}_{-1.45}   $& 
			$5.53 ^{+1.62}_{-1.52}   $& 
			$5.22 ^{+1.56}_{-1.50} $& 
			$5.23 ^{+1.54}_{-1.58}$\\
			
			{\boldmath$10^3\Omega_K$} & 
			$1.55^{+4.91}_{-4.54} $   & 
			$0.80^{+4.72}_{-4.53} $  & 			$5.10^{+5.97}_{-5.94}$   & 
			$4.53^{+5.92}_{-5.99}$\\
			
			{\boldmath$\Sigma m_\nu $   [eV] } & 
			$<0.199$& 
			$<0.188$& 
			$0.256^{+0.224}_{-0.234}$& 
			$0.257^{+0.219}_{-0.234}$\\
			
			{\boldmath$10^4 \nu          $}&
			$<1.36$&
			$-$&
			$<1.39$&
			$-$\\
			
			$H_0 $            [km/s/Mpc]            & $67.8^{+1.4}_{-1.3}           $& 
			$67.9^{+1.3}_{-1.2}           $& 
			$67.9^ \pm 1.4                $& 
			$68.1^{+1.4}_{-1.3}           $\\
			
			$\sigma_8                         $& $0.808^{+0.027}_{-0.034}  $& $0.810^{+0.026}_{-0.031}  $& 
			$0.764^{+0.040}_{-0.042}   $&
			$0.765\pm 0.040  $\\

			\hline
			$\chi^2_{best-fit}$& 
			$3472.32   $&
			$3474.92   $& 
			$3523.74   $&
			$3524.51   $\\
			\hline
		\end{tabular}
		\label{tab:fit}
	\end{center}
\end{table}

In order to constrain the cosmological parameters of RVM and $\Lambda$CDM in the non-flat universe, we use the {\bf CosmoMC} package with a MCMC engine to explore the parameter space with the combinations of the observational data sets, which include the CMB temperature fluctuation from
 {\it Planck 2018} 
with TT, TE, EE, low-$l$ polarization from SMICA~\cite{Aghanim:2019ame, Aghanim:2018eyx,Aghanim:2018oex,Akrami:2019izv}, 
BAO data from 6dF Galaxy Survey~\cite{Beutler:2011hx} and BOSS~\cite{Anderson:2013zyy}, { supernova(SN) data from the JLA compilation~\cite{Betoule:2014frx}, } the weak lensing (WL) data from CFHTLenS ~\cite{Heymans:2013fya} and direct large scale structure (LSS) formation data, and the data points of $f\sigma_8$ listed in Table~\ref{tab:2}. The priors of parameters are given in Table~\ref{tab:prior}. Due to the tension between the geometry data (SNIa, BAO etc.) and growth data (WL, $f\sigma_8$) ~\cite{Lin:2017ikq}, we choose the two combinations of CMB+BAO+SN and CMB+BAO+SN+WL+$f\sigma_8$ in our fits. To calculate the best fitted values of $\chi^2$, we use that
\begin{eqnarray}
\chi^2_c = \sum_{i=1}^n \frac{(T_c(z_i) - O_c(z_i))^2}{E^i_c} \,,
\end{eqnarray}
where $c$ denotes the type of the data, $n$ is the number of the data in each data set, $T_c$ represents the theoretical value derived form {\bf CAMB} at 
the redshift $z_i$, and $O_c$ ($E_c$) corresponds to the observational value (covariance).

The global fitting results of RVM and $\Lambda$CDM in the non-flat universe are plotted in Figs.~\ref{fig:2018cmbbaosn} 
and \ref{fig:2018cmbbaosnwlfsigma8}, while those listed in Table~\ref{tab:fit} correspond to the cosmological parameters and $\nu$, given at 95$\%$ and 68$\%$ C.L., respectively. Our results show that $\nu\lesssim 1.39 \times10^{-4}$ at 68 $\%$C.L. in the non-flat universe of RVM for the data set of CMB+BAO+SN+WL+$f\sigma_8$, which is similar to the previous result of $1.54\times10^{-4}$ at 68 $\%$ C.L. in RVM for the flat universe~\cite{Zhang2019}. 
Explicitly, we obtain that $\chi^2_{RVM}=3472.32~(3523.74)$ and $\chi^2_{\Lambda CDM}=3474.92~(3524.51)$
when fitting with the data set of CMB+BAO+SN (CMB+BAO+SN+WL+$f\sigma_8$),
indicating that our results in RVM are consistent with those in $\Lambda$CDM for the  non-flat universe. 
For the density parameter $\Omega_K$ of the spatial curvature at the present time, our results show that an open universe is preferred instead of the closed one in Ref.~\cite{Valentino2019}.
 In particular,
 when the data of WL and $f\sigma_8$ are included, both RVM and $\Lambda$CDM favor the open universe
 with $|\Omega_K|\leq O(10^{-2})$. 
Our result of the open universe is consistent with that in Ref.~\cite{Farrugia2018}.


On the other hand, the best-fit values of the neutrino mass sum, $\Sigma m_{\nu}$, 
in the non-flat universe are similar with those in the flat universe 
when fitting with CMB+BAO+SN~\cite{Geng20172,Giusarma2016,Giusarma2018,Vagnozzi2017}.
However, it is interesting to see that the constraints on $\Sigma m_{\nu}$ are relaxed
for the data sets of CMB+BAO+SN+WL+$f\sigma_8$,
in which the $f\sigma_8$ data points play the main role.
Note that similar results are also obtained  in the flat universe as shown in Ref.~\cite{Geng20172}.
This is because the structure-growth rate of $f\sigma_8$ is a unique indicator of massive
neutrinos~\cite{Boyle:2017lzt,Boyle:2018rva}.
Specifically, we have $\Sigma m_{\nu}=0.256^{+0.224}_{-0.234}$ ~($0.257^{+0.219}_{-0.234}$)~eV at 95 $\%$C.L., resulting in the non-zero lower bounds of $\Sigma m_{\nu} \geq 0.022~(0.023)$ eV at 95$ \%$ C.L. for RVM ($\Lambda$CDM) in the spatially curved universe.
On the other hand,  due to the possible degeneracy among $\tau$, $\Omega_K$ and $\Sigma m_{\nu}$,
the constraints on $\tau$ and $\Omega_K$ are not improved  with the data points of WL+$f\sigma_8$  as seen from Table~\ref{tab:fit}.
To obtain better constraints, one would use some additional data, such as  $21~cm$ emission measurements~\cite{21cm,Archidiacono:2016lnv}, 
which could fix the parameter $\tau$ to break the degeneracy of $\tau$, $\Omega_K$ and $\Sigma m_{\nu}$~\cite{Boyle:2017lzt,Boyle:2018rva}.
We note that in our data fitting we do not specify the neutrino mass hierarchy  in  $\Sigma m_\nu$. 
For the cosmological effects of the  neutrino mass hierarchy, one can refer to  
the discussions in the literature~\cite{Xu20161, Xu20162, Huang20161}.

\section{Conclusions}\label{sec:conclusion}
We have studied the model with the running cosmological constant of $\Lambda=3\nu (H^{2}+ K/a^2)+c_0$ in 
the spatially curved universe. 
We have compared our results  for several sets of $\nu$ and $\Omega_K$ with the Planck 2018 data
in the CMB power spectra. 
We have found that $\nu$ and $\Omega_K$ have similar effects on 
the CMB power spectra, but only non-zero values of $\nu$ would lead to large suppressions in the CMB TT mode spectra. 
In the two combinations of the observational data, we have constrained that $\nu \leq O(10^{-4})$ together with $|\Omega_K|\leq O(10^{-2})$. 
Notably, the constraints on $\nu$ in the non-flat universe are similar to those in the flat universe. 
From the best fitted values of $\chi^2$, we have shown that RVM is in consistent with $\Lambda$CDM. 
When fitting with the date set of CMB+BAO+SN+WL+$f\sigma_8$, we have obtained the non-zero lower bounds of $\Sigma m_{\nu} \geq 0.022$ and $0.023$ eV at 95$\%$ C.L. in the non-flat
RVM and $\Lambda$CDM, respectively,
  indicating that the involvement of a non-zero $\Omega_K$ would provide  viable constraints on the absolute 
neutrino masses in cosmological models.

\acknowledgments  
The work was supported in part by National Center for Theoretical Sciences, MoST (MoST-107-2119-M-007-013-MY3), National Natural Science Foundation of China under Grants No. 11505004 and No. 11447104,  and the Anhui Provincial Natural Science Foundation of China under Grant No. 1508085QA17. K. Z. would like to thank Dr. Yangsheng Yuan and Shandong Normal University for part of numerical calculations run by the workstation of their group. 


\end{document}